\documentclass[12pt]{article}
\usepackage{epsf,graphicx,amsmath,psfrag}

\def\bm#1{\mbox{\boldmath$#1$\unboldmath}}
\newcommand\mg{m_{\tilde g}}
\newcommand\mb{m_{\tilde b}}

\begin{document}

\begin{titlepage}

\begin{flushright}
CLNS~02/1790\\
SLAC-PUB-9230\\
{\tt hep-ph/0205274}\\[0.2cm]
May 24, 2002
\end{flushright}

\vspace{1.0cm}
\begin{center}
\Large\bf 
Constraints on Light Bottom Squarks from\\ 
Radiative B-Meson Decays
\end{center}

\vspace{0.5cm}
\begin{center}
Thomas Becher \\[0.1cm]
{\sl Stanford Linear Accelerator Center\\
Stanford University, Stanford, CA 94309, USA}\\[0.5cm]
Stephan Braig, Matthias Neubert\\[0.1cm]
{\sl Newman Laboratory of Elementary-Particle Physics, Cornell University\\
Ithaca, NY 14853, USA}\\[0.5cm]
Alexander L.\ Kagan\footnote{On leave from: Department of Physics, 
University of Cincinnati, Cincinnati, Ohio 45221, USA}\\[0.1cm]
{\sl Fermi National Accelerator Laboratory\\
Batavia, IL 60510, USA}
\end{center}

\vspace{1.0cm}
\begin{abstract}
\vspace{0.2cm}\noindent 
The presence of a light $\tilde b$ squark ($\mb\sim 4$\,GeV) and gluino 
($\mg\sim 15$\,GeV) might explain the observed excess in $b$-quark 
production at the Tevatron. Though provocative, this model is not 
excluded by present data. The light supersymmetric particles can 
induce large flavor-changing effects in radiative decays of $B$ mesons.
We analyse the decays $B\to X_s\gamma$ and $B\to X_{sg}$ in this
scenario and derive restrictive bounds on the flavor-changing 
quark-squark-gluino couplings.
\end{abstract}
\vfill

\end{titlepage}

\section{Motivation}

The measured $b$-quark production cross section at hadron colliders 
exceeds next-to-leading order (NLO) QCD predictions by more than a factor 
of two. While it is conceivable that this discrepancy is due to 
higher-order corrections, the disagreement is surprising since NLO 
calculations have been reliable for other processes in this energy range. 
Berger et al.\ have analysed $b$-quark production in the context of
the Minimal Supersymmetric Standard Model (MSSM) and find that the excess in 
the cross section could be attributed to gluino pair-production followed by 
gluino decay into pairs of $b$ quarks and $\tilde b$ squarks, if both the 
gluino and the $\tilde b$ squark are sufficiently light 
\cite{Berger:2000mp}. In order to reproduce the transverse-momentum 
distribution of the $b$ quarks, the masses of the gluino and light 
$\tilde b$-squark mass eigenstate should be in the range 
$\mg=12$--16\,GeV and $\mb=2$--5.5\,GeV. The masses of all other 
supersymmetric (SUSY) particles are assumed to be large, of order several
hundred GeV, so as to have evaded detection at LEP2. Interestingly, a 
renormalization-group analysis in the framework of the unconstrained MSSM 
shows that a light $\tilde b$ squark is most natural if it is accompanied by 
a light gluino with mass of order 10\,GeV \cite{Dedes:2000nv}.

Berger et al.\ have further observed that a light $\tilde b$ squark could 
have escaped direct detection. For example, the additional contribution to 
the $e^+ e^-\to\text{hadrons}$ cross section at large energy would only be 
about 2\% and hence difficult to disentangle. The pair-production of light 
scalars would alter the angular distribution of hadronic jets in $e^+ e^-$ 
collisions, but the present data are not sufficiently precise to rule out 
the existence of this effect \cite{Berger:2000mp}. On the other hand, there 
are important $Z$-pole constraints on the parameters of this model. Most 
importantly, production of the light $\tilde b$ squark at the $Z$ pole has 
to be suppressed, which implies a stringent constraint on the mixing angle 
$\theta$ relating the sbottom mass and weak eigenstates \cite{Carena:2001ka}.
More recently, several authors have studied loop effects of the light
SUSY particles on electroweak precision measurements 
\cite{Cao:2001rz,Cho:2002mt,Baek:2002xf}, finding potentially large 
contributions to the quantity $R_b$. However, a conflict with existing data 
can be avoided by having some of the superpartner masses near current 
experimental bounds, or by allowing for a new CP-violating phase in the SUSY 
sector \cite{Baek:2002xf}.

The null result of a CLEO search for the semileptonic decays 
$\tilde B\to D^{(*)} l\pi$ and $\tilde B\to D^{(*)} l\tilde\chi^0$ of 
sbottom hadrons implies that the branching ratios for the decays 
$\tilde b\to c\,l$ induced by $R$-parity violating couplings, or 
$\tilde b\to c\,l\chi^0$ with an ultra-light neutralino $\chi^0$, must be 
highly suppressed \cite{Savinov:2000jm}. However, a light $\tilde b$ 
squark would be allowed to decay promptly via hadronic $R$-parity 
violating couplings in the modes $\tilde b\to\bar c\bar q$ or 
$\tilde b\to\bar u\bar q$ (with $q=u,s$). Alternatively, it could be 
long-lived, forming $\tilde b$-hadrons. An interesting consequence of 
hadronic $R$-parity violating decays would be the abundant production of 
light baryons. This could significantly alter the thrust-axis angular 
distribution for continuum events containing baryons at the $B$ factories.

A striking manifestation of the light $\tilde b$-squark scenario 
would be the production of like-sign charged $B$ mesons at hadron colliders, 
because the Majorana nature of the gluino allows for the production of 
$bb\tilde b^*\tilde b^*$ and $\bar b\bar b\tilde b\tilde b$ final states
\cite{Berger:2000mp}. Another potential signature at hadron colliders is an 
enhanced yield of $t\bar{t} b \bar{b}$ events \cite{Leibovich:2002qp}.  
It has also been pointed out that sbottom pairs 
would be copiously produced in $\Upsilon(nS)\to\tilde b\tilde b^*$ and 
$\chi_{bJ}\to\tilde b\tilde b^*$ decays \cite{Berger:2001jb,Berger:2002gu}. 
Precise measurements of bottomonium decays could lead to new constraints 
on the squark and gluino masses.

The presence of light SUSY particles alters the running of $\alpha_s$, and 
it is often argued that this would exclude the existence of light gluinos. 
This argument is incorrect. First, a gluino with mass $\mg\sim 15$\,GeV 
would have a relatively small effect on the evolution of $\alpha_s$. Taking, 
for instance, $\alpha_s(m_b)=0.21$ (a value in agreement with all low-energy 
determinations of the QCD coupling) and including the contribution of the 
gluino octet to the $\beta$ function above the scale $\mu_{\tilde g}=\mg$
yields $\alpha_s(m_Z)=0.126$, which is about three standard deviations 
higher than the canonical value $\alpha_s(m_Z)=0.118\pm 0.003$. However, 
considering that at leading order only virtual gluino
pairs contribute to the $\beta$ function, a more realistic treatment would
include the gluino contribution above a scale $\mu_{\tilde g}=2\mg\sim 
30$\,GeV, in which case $\alpha_s(m_Z)=0.121$, in good agreement with the 
standard value. Secondly, it is important to realize that even a value of
$\alpha_s(m_Z)$ significantly above 0.118 would not rule out the model,
the reason being that the characteristic scale $\mu$ inherent in all 
determinations of $\alpha_s(\mu)$ is typically much smaller than the total
energy. This is true, in particular, for the determinations based on 
event-shape variables. In practice, the measurements fix 
$\alpha_s(\mu)$ somewhere between a fraction of the $Z$ mass down to several 
GeV, where the gluino contribution to the $\beta$ function is negligible.
Using these determinations to quote values of $\alpha_s(m_Z)$ (as is 
routinely done) assumes implicitly that the coupling runs as predicted in
the SM. Finally, a careful analysis of the running of $\alpha_s$ in the
presence of light SUSY particles would have to include, for each observable,
the modifications in the theoretical formulae due to virtual and real 
emissions of the new particles. These corrections could be significant, and 
could partially compensate effects arising from the modification of the 
$\beta$ function.

If we are to take the possibility of a light $\tilde b$ squark and light 
gluinos seriously, then the theoretical study of their impact must be 
extended to the phenomenology of weak decays of the $b$ quark. New sources 
of flavor violation arise from $s$-$\tilde b$-$\tilde g$ and 
$d$-$\tilde b$-$\tilde g$ couplings. The overall scale of SUSY 
flavor-changing interactions originating from gluino exchange is set by 
the factor $g_s^2/\mg^2$, which is much larger than the corresponding factor 
$G_F\sim g_W^2/m_W^2$ for weak decays in the Standard Model (SM).
Consequently, the new flavor-changing couplings must be much smaller than 
the CKM mixing angles in order for this model to be 
phenomenologically viable. The most stringent bounds arise from the 
radiative decay $B\to X_s\gamma$, which we discuss in the present work.
(Contributions of light $\tilde b$ squarks to kaon decays, $K$-$\bar K$
mixing, and $D$-$\bar D$ mixing are strongly suppressed.) The presence of
such tight bounds implies stringent constraints on model building.   

If the light $\tilde b$ squark is sufficiently light to be pair-produced in 
$b$ decays, new unconventional decay channels would be opened up, which 
could affect the phenomenology of 
$B$ mesons and beauty baryons. Examples of potentially interesting 
consequences include modifications of beauty lifetime ratios, an 
enhancement of the semileptonic branching ratio of $B$ mesons via production
of charmless final states containing $\tilde b$ squarks, an enhancement of 
$\Delta\Gamma(B)$ and of the semileptonic CP asymmetry $A_{\rm SL}$, and 
wrong-sign kaon production via $b\to\bar s\tilde b\tilde b$ transitions
allowed by the Majorana nature of the gluino. The phenomenology of such 
effects will be discussed elsewhere. If the light $\tilde b$ squark is too 
heavy to be pair-produced, it would still give rise to potentially large 
virtual effects in $B$ decays. Their study is the main purpose of this 
Letter.

\section{\boldmath 
The low-energy effective $\Delta B=1$ Hamiltonian\unboldmath}

We denote by $\tilde d_i$ with $i=1,..,6$ the down-squark mass eigenstates, 
and by $\tilde q_L$ and $\tilde q_R$ with $q=d,s,b$ the interaction 
eigenstates (the superpartners of the left-handed and right-handed down 
quarks). They are related by a unitary transformation 
$\tilde q_L=\Gamma_{qi}^{L\dagger}\,\tilde d_i$ and
$\tilde q_R=\Gamma_{qi}^{R\dagger}\,\tilde d_i$. We identify $\tilde d_3$ 
with the light sbottom mass eigenstate and define a sbottom-sector mixing 
angle $\theta$ through $\Gamma_{b3}^R=\cos\theta$ and 
$\Gamma_{b3}^L=\sin\theta$. The fact that the light $\tilde b$ squarks are 
not produced in $Z$ decays implies
\begin{equation}
   \sin\theta \approx \pm\sqrt{\frac23}\,\sin\theta_W 
   \approx \pm\sqrt{\frac23\left( 1 - \frac{m_W^2}{m_Z^2}\right)} \,.
\end{equation}
The phenomenologically favored range for the $Z\tilde d_3\tilde d_3$ 
coupling is $|\sin\theta|=0.3$--0.45 \cite{Carena:2001ka}, meaning that the 
light sbottom is predominantly the superpartner of the right-handed bottom 
quark. In our numerical analysis we will assume a vanishing tree-level 
coupling to the $Z$ and thus use $\sin\theta=\pm 0.395$. 

The flavor-changing couplings involving the light $\tilde b$ and $\tilde g$ 
fields can be parameterized by dimensionless quantities 
\begin{equation}
   \epsilon_{qb}^{AB} = \Gamma_{q3}^{A\dagger}\,\Gamma_{b3}^B \,, \qquad
   (\text{with} ~\epsilon_{qb}^{AL} = \epsilon_{qb}^{AR}\,\tan\theta)
\end{equation}
where $A,B=L,R$, and $q=s$ or $d$ for $b\to s$ or $b\to d$ transitions,
respectively. In general the parameters $\epsilon_{qb}^{AB}$ are 
complex, which can lead to new CP-violating effects. These parameters are 
invariant under a phase redefinition of the light $\tilde b$-squark state, 
and they transform in the same way as the products $V_{iq}^* V_{ib}$ (with 
$i=u,c,t$) of CKM matrix elements under a phase redefinition of the 
down-type quark fields. It follows that ratios of the type 
$\epsilon_{qb}^{AB}/\epsilon_{qb}^{CD}$ and 
$\epsilon_{qb}^{AB}/(V_{iq}^* V_{ib})$ are invariant under phase 
redefinitions, and thus can carry an observable, CP-violating phase.

\begin{figure}
\begin{center}
\includegraphics[width=0.6\textwidth]{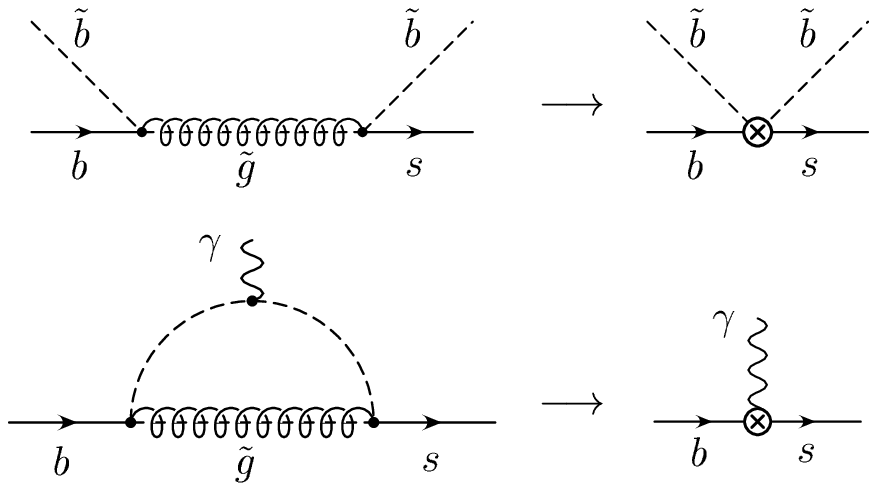}
\vspace*{-0.2cm}
\parbox{15cm}{\caption{\label{fig:effTh}
Examples of $b\to s$ transitions induced by gluino exchange. The diagrams 
on the right show the corresponding contributions in the effective theory 
where the gluinos are integrated out.}}
\end{center}
\end{figure}

Flavor-changing hadronic processes in the model with a light gluino and a 
very light $\tilde b$ squark are most transparently described by means of 
an effective ``weak'' Hamiltonian. If we neglect effects that are 
suppressed by inverse powers of the heavy SUSY scale, the relevant energy 
scales are the electroweak scale, at which the usual SM flavor-changing 
operators are generated by integrating out the top quark and the $W$ and 
$Z$ bosons, and the scale $\mg$, at which new flavor-changing operators 
are generated by integrating out the gluinos. We start by discussing the 
construction of the effective theory below the gluino scale, focusing on 
the new interactions proportional to $\epsilon_{qb}^{AB}$ induced by gluino 
exchange, as illustrated in Figure~\ref{fig:effTh}. SUSY modifications of 
the renormalization-group (RG) evolution of the standard weak-interaction 
operators will be discussed later. The remaining light degrees of freedom 
in the low-energy theory are the quarks $u,d,s,c,b$, the photon and gluons, 
and the light $\tilde b$ squark. Operators in the effective Hamiltonian can 
be organized in an expansion in inverse powers of the gluino mass. For 
$\mg\approx 15\,\text{GeV}\gg m_b$, it is a good approximation to keep the 
leading terms in this expansion, which have mass dimension five. These 
operators comprise the usual electromagnetic and chromomagnetic dipole 
operators, and new operators containing two scalar $\tilde b$ fields. The 
effective Hamiltonian for $b\to s$ transitions is (here and below, 
$\mg\equiv\mg(\mg)$ denotes the running gluino mass at the gluino matching 
scale)
\begin{equation}
   H_{\rm eff}^{\rm SUSY}
   = \frac{4\pi\alpha_s(\mg)}{\mg}\,
   \sum_i {\cal C}_i(\mu)\,\Big[ \epsilon_{sb}^{LR}\,O_i^{LR}(\mu)
   + (L\leftrightarrow R) \Big] + O(1/\mg^2) \,,
\end{equation}
where
\begin{align}
   O_1^{LR} &= \bar s_L t_a\tilde b\,\tilde b^* t_a b_R \,, &
   O_2^{LR} &= \bar s_L \tilde b\,\tilde b^* b_R \,, \nonumber\\
   O_7^{LR} &= -\frac{e}{16\pi^2}\,
    \bar s_L\sigma_{\mu\nu} F^{\mu\nu} b_R \,, &
   O_8^{LR} &= -\frac{g_s}{16\pi^2}\,
    \bar s_L\sigma_{\mu\nu} G^{\mu\nu} b_R \,. 
\end{align}
We work with the covariant derivative
$iD^\mu=i\partial^\mu+e Q_d A^\mu+g_s A_a^\mu t_a$, where $Q_d=-\frac13$ is 
the electric charge of a down-type (s)quark. (To facilitate comparison 
with the literature, which usually adopts the opposite sign convention for 
the couplings, we have included a factor of $-1$ in the definition of the 
dipole operators $O_7$ and $O_8$.) In addition, there are dimension-five 
fermion-number violating interactions of the form 
$\bar s^c(1\pm \gamma_5 )b\,\tilde b^*\tilde b^*$, which mediate 
$b\to\bar s\tilde b\tilde b$ transitions. They are irrelevant to our 
discussion here.

\begin{figure}
\begin{center}
\begin{tabular}{ccc}
\includegraphics[width=0.27\textwidth]{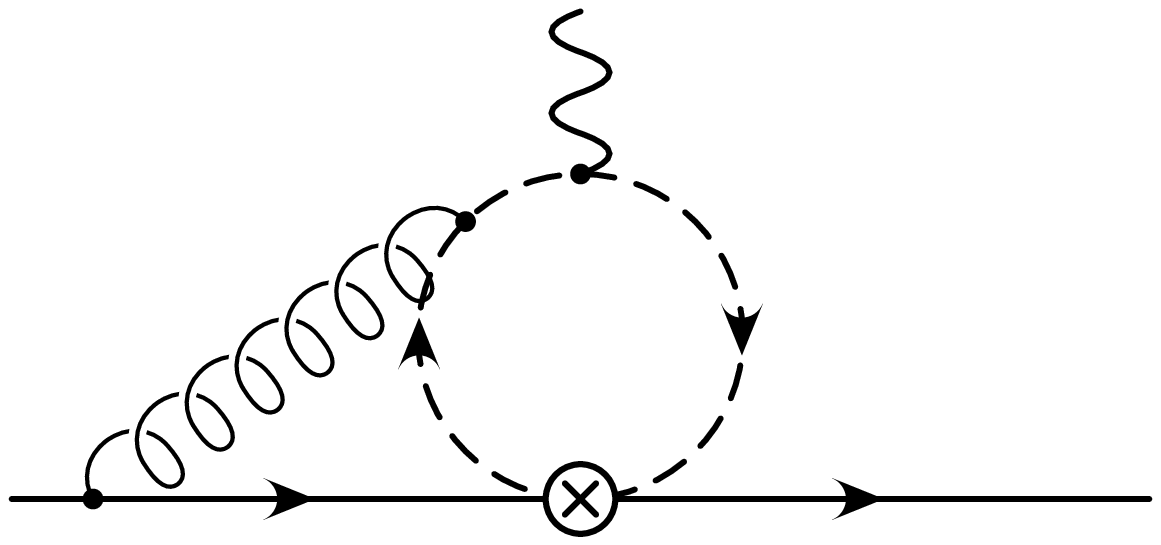} &
\includegraphics[width=0.27\textwidth]{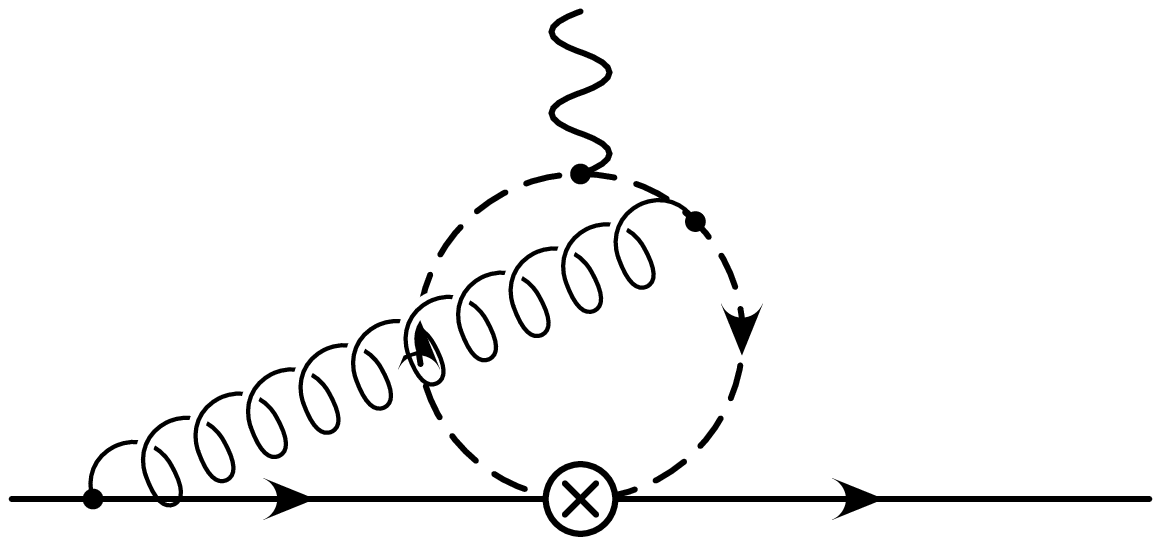} &
\includegraphics[width=0.27\textwidth]{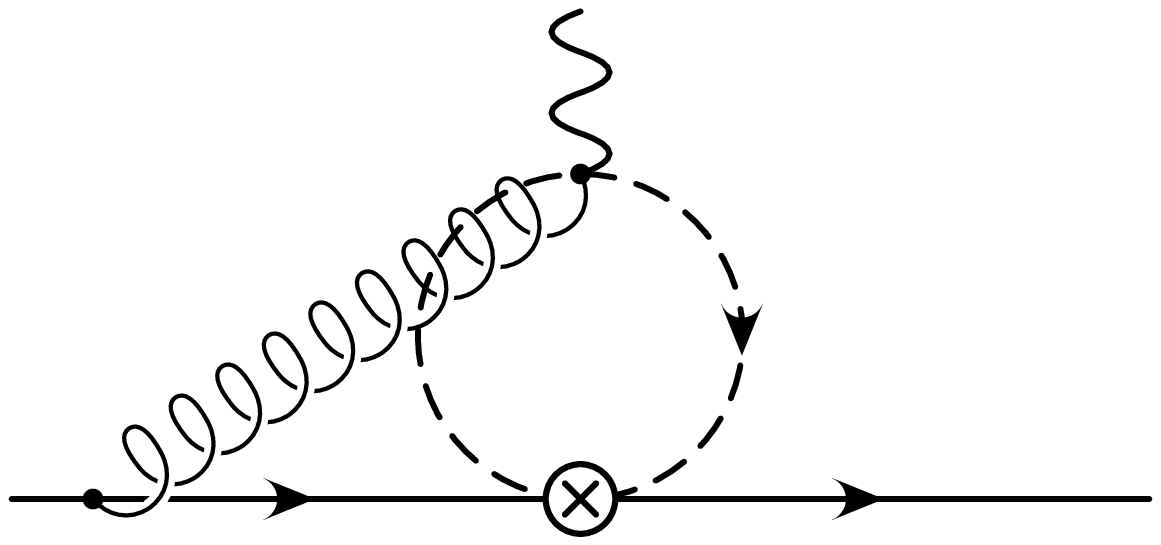}\\
 $D_1$ & $D_2$ & $D_3$
\end{tabular}
\end{center}
\vspace*{-0.5cm}
\begin{center}
\begin{tabular}{|l|c|cc|cc|}
\hline\hline
 & & $O_1\to O_7$ & $O_2\to O_7$ & $O_1\to O_8$ & $O_2\to O_8$ \\ 
\hline
$D_1$ & $-\frac{1}{8\epsilon^2}+\frac{3}{16\epsilon}$
 & $Q_d\,(-\frac14+\frac{1}{4N^2})$ & $Q_d\,(\frac{N}{2}-\frac{1}{2N})$
 & $\frac{1}{4N^2}$ & $\frac{N}{2}-\frac{1}{2N}$ \\
$D_2$ & $-\frac{1}{8\epsilon^2}+\frac{3}{16\epsilon}$
 & $Q_d\,(-\frac14+\frac{1}{4N^2})$ & $Q_d\,(\frac{N}{2}-\frac{1}{2N})$
 & $\frac14 + \frac{1}{4N^2}$ & $-\frac{1}{2N}$ \\
$D_3$ & $\phantom{-}\frac{1}{4\epsilon^2}-\frac{1}{8\epsilon}\,\,$
 & $Q_d\,(-\frac14+\frac{1}{4N^2})$ & $Q_d\,(\frac{N}{2}-\frac{1}{2N})$
 & $\frac18+\frac{1}{4N^2}$ & $\frac{N}{4}-\frac{1}{2N}$ \\
\hline\hline
\end{tabular}
\end{center}
\vspace*{-0.2cm}
\centerline{\parbox{15cm}{\caption{\label{fig:mixing}
Two-loop diagrams relevant to the mixing of $O_{1,2}$ into $O_{7,8}$, and
corresponding results, in units of $\alpha_s/4\pi$, after the subtraction 
of subdivergences. Mirror-symmetric graphs with the gluon attached to the 
$s$-quark line give identical contributions. Results in the first column 
of the table have to be multiplied by the color and charge factors in the 
remaining columns.}}}
\end{figure}

The Wilson coefficients at a scale $\mu\sim\mg$ are obtained by matching 
the effective theory to the full theory. At leading order we find
\begin{equation}
   {\cal C}_1(\mg) = 2 \,, \quad
   {\cal C}_2(\mg) =0 \,, \quad
   {\cal C}_7(\mg) = Q_d\,\frac{N^2-1}{4N} \,, \quad
   {\cal C}_8(\mg) = - \frac{N^2+1}{4N} \,,
\end{equation}
where $N=3$ is the number of colors. In order to use the effective 
Hamiltonian for calculating $B$-decay amplitudes, we compute the values of 
the Wilson coefficients at a low scale $\mu\sim m_b$ by solving the RG 
equation $(d/d\ln\mu-\bm{\gamma}^T)\,\vec{C}(\mu)=0$. At leading order, the 
anomalous dimension matrix $\bm{\gamma}$ 
receives contributions from the one-loop mixing 
of the operators $(O_1,O_2)$ and $(O_7,O_8)$ among themselves, and from the
two-loop mixing of $O_{1,2}$ into $O_{7,8}$. This is analogous to the case 
of the SM, in which one needs to consider the two-loop mixing of the 
current-current operators into the dipole operators at leading order 
\cite{Ciuchini:1993ks}. In our case, only the three two-loop diagrams shown 
in Figure~\ref{fig:mixing} give a nonvanishing contribution. All other 
graphs vanish after their subdivergences are removed. The calculation of 
the UV divergences of these diagrams can be reduced to the evaluation of 
massive tadpole integrals \cite{Chetyrkin:1997fm}. The resulting anomalous 
dimension matrix in the operator basis $(O_1,O_2,O_7,O_8)$ reads
\begin{equation}
   \bm{\gamma} = \frac{\alpha_s}{4\pi} \left( 
   {\arraycolsep=0.3cm\begin{matrix}
    -6N + \frac{9}{N} & -\frac32 + \frac{3}{2N^2}
     & Q_d\,(\frac14 - \frac{1}{4N^2}) &  -\frac18 - \frac{1}{4N^2} \\ 
    -6 & -3N + \frac{3}{N} & Q_d\,(-\frac{N}{2} + \frac{1}{2N})
     & -\frac{N}{4} + \frac{1}{2N} \\ 
    0 & 0 & N - \frac{1}{N} & 0 \\ 
    0 & 0 & Q_d\,(4N - \frac{4}{N}) & N - \frac{5}{N} \\  
   \end{matrix}} \right) + O(\alpha_s^2) \,.
\end{equation}

The scale dependence of the Wilson coefficients is now readily obtained by 
solving the RG equation. Setting $N=3$, we find
\begin{eqnarray}
   {\cal C}_1(\mu) &=& \frac{16}{9}\,\eta^{-8} + \frac29\,\eta^{-7/2} \,,
    \qquad
    {\cal C}_2(\mu) = \frac{8}{27} \left( \eta^{-8} - \eta^{-7/2}
    \right) , \nonumber\\
   {\cal C}_7(\mu) &=& Q_d \left( -\frac{4}{273}\,\eta^{-8}
    - \frac{4}{145}\,\eta^{-7/2} + \frac{438}{65}\,\eta^{2/3}
    - \frac{1224}{203}\,\eta^{4/3} \right) , \nonumber\\
   {\cal C}_8(\mu) &=& \left( \frac{1}{39}\,\eta^{-8}
    - \frac{1}{60}\,\eta^{-7/2} - \frac{219}{260}\,\eta^{2/3} \right) .
\end{eqnarray}
Here $\eta=[\alpha_s(\mg)/\alpha_s(\mu)]^{1/\beta_0(5,0,1)}$, and
\begin{equation}\label{beta0}
   \beta_0(n_f,n_g,n_s) = \frac{11N}{3} - \frac23\,n_f
   - \frac{2N}{3}\,n_g - \frac16\,n_s
\end{equation}
is the first coefficient of the generalized QCD $\beta$ function in the 
presence of $n_f$ light Dirac fermions, $n_g$ light gluino octets, and 
$n_s$ light complex scalars. Numerical results for the coefficients 
${\cal C}_i(\mu)$ will be given in Table~\ref{tab:Ccoefs} below. The scale 
dependence of the Wilson coefficients below the scale $\mg$ arises mainly 
from the mixing of $O_1$ with $O_2$ and $O_7$ with $O_8$. The mixing of 
$O_1$ and $O_2$ into the dipole operators turns out to be small numerically. 

The presence of light SUSY particles also affects the RG evolution of the 
SM contributions to the effective weak Hamiltonian below the electroweak 
scale. We will now discuss these effects for the operators of relevance to 
radiative $B$ decays.

\section{\boldmath 
The radiative decay $B\to X_s\gamma$\unboldmath}

The inclusive radiative decay $B\to X_s\gamma$ is one of the most 
sensitive probes of physics beyond the SM. Indeed, we will see that this 
decay provides very stringent bounds on the flavor-changing couplings 
$\epsilon_{sb}^{LR}$ and $\epsilon_{sb}^{RL}$. The SM 
prediction for the $B\to X_s\gamma$ decay rate is known at NLO 
\cite{Greub:1996tg,Chetyrkin:1996vx,Kagan:1998ym,Gambino:2001ew} and, 
within errors, agrees with the data. The change of this prediction due to 
the light SUSY particles present in our model is fourfold:

\begin{enumerate}
\item
The main effects are the genuine SUSY flavor-changing interactions due to
quark-squark-gluino couplings. For $\mu<\mg$, these interactions are 
described by the effective Hamiltonian constructed in the previous section.
\item
Even in the absence of flavor-changing couplings in the SUSY sector (i.e., 
for vanishing $\epsilon_{qb}^{AB}$), the SM operator basis gets enlarged 
by dimension-six penguin operators with field content 
$\bar s b\,\tilde b^* i\!\overleftrightarrow{D}^\mu\tilde b$ and 
$\bar s b\,\bar{\tilde g}\tilde g$. 
These new operators have small Wilson coefficients and yield a
negligible contribution to the $B\to X_s\gamma$ decay rate. In our 
analysis, we will neglect these as well as all four-quark penguin 
operators.
\item
The presence of a light gluino octet (and, to lesser extent, of a 
light $\tilde b$ scalar) modifies the running of the strong coupling 
constant. We use the two-loop expression for $\alpha_s(\mu)$, modified to 
account for the effects of the light SUSY particles.
\item
There is a SUSY contribution to the $b$-quark wave-function renormalization,
which adds to the anomalous dimensions of the SM operators. Also, the 
masses of the $b$ quark and the gluino mix under renormalization, thus 
altering the scale dependence of $m_b(\mu)$. This is a novel effect due to 
the decoupling of the heavy $\tilde b_H$ squark at the SUSY scale. 
\end{enumerate}

\noindent
The last two effects change the evolution of the Wilson coefficients of the 
SM operators between the electroweak scale and the scale $\mu\sim\mg$, where
the gluino degrees of freedom are integrated out. (Beyond leading order, 
the anomalous dimensions of the SM operators are also changed due to 
internal loops involving SUSY particles.) 

The effective weak Hamiltonian governing $B\to X_s\gamma$ decays in the SM
is 
\begin{equation}\label{Heff}
   H_{\rm eff}^{\rm weak} = -\frac{4 G_F}{\sqrt2}\,V_{ts}^* V_{tb}
   \sum_i C_i(\mu_b)\, Q_i(\mu_b) \,.
\end{equation}
The operators relevant to our calculation are
\begin{equation}
   Q_1 = \bar s_L^i\gamma_\mu c_L^j\,\bar c_L^j\gamma^\mu b_L^i \,, \quad
   Q_2 = \bar s_L\gamma_\mu c_L\,\bar c_L\gamma^\mu b_L \,, \quad
   Q_{7\gamma} = m_b O_7^{LR} \,, \quad 
   Q_{8g} = m_b O_8^{LR} \,.
\end{equation}
To an excellent approximation, the contributions of other operators can be 
neglected. To obtain the values of the corresponding Wilson coefficients 
$C_i(\mu_b)$ in our model, we first evolve them from the electroweak scale 
$\mu=m_W$ down to a scale $\mu_{\tilde g}\sim \mg$, and in a second step 
from $\mu_{\tilde g}$ to a scale $\mu_b\sim m_b$. 

Above the gluino scale, there are SUSY contributions to the 
wave-function renormalization constants of left- and right-handed $b$-quark 
fields from gluino-squark loops. At one-loop order, we obtain the 
gauge-independent results
\begin{equation}\label{Zlr}
   \delta Z_2(b_L) = - \frac{C_F\alpha_s}{4\pi\epsilon}\,\sin^2\theta
   \,, \qquad
   \delta Z_2(b_R) = - \frac{C_F\alpha_s}{4\pi\epsilon}\,\cos^2\theta \,.
\end{equation}
Next, by calculating the self-energies of $b$ quarks and gluinos we find 
that their masses mix under renormalization. The corresponding anomalous 
dimension matrix in the basis $(m_b,\mg)$, defined such that 
$dm_i/d\ln\mu=-(\bm{\gamma}_m)_{ij}\,m_j$, reads
\begin{equation}\label{massmix}
   \bm{\gamma}_m = \frac{\alpha_s}{4\pi} \left( 
   {\arraycolsep=0.3cm\begin{matrix}
    \frac{5N}{2} - \frac{5}{2N} & \left(N-\frac{1}{N}\right) \sin 2\theta \\
    \sin 2\theta & 6N - \frac12
   \end{matrix}} \right) + O(\alpha_s^2) \,.
\end{equation}
Note that the off-diagonal entries are sensitive to the sign of the mixing
angle $\theta$. The RG evolution of the operators $Q_i$ in (\ref{Heff}) is 
complicated by the effects of mass mixing. To compute the resulting 
modifications of the Wilson coefficients we work in the extended operator 
basis $(Q_1,Q_2,m_b O_7^{LR},m_b O_8^{LR},\mg O_7^{LR},\mg O_8^{LR})$. 
Using (\ref{Zlr}), (\ref{massmix}), and the well-known anomalous dimensions 
of the SM operators \cite{Ciuchini:1993ks}, we obtain for the anomalous 
dimension matrix $\bm{\gamma}_Q=\frac{\alpha_s}{4\pi}\,\bm{\gamma}_Q^{(0)}
+O(\alpha_s^2)$, where (setting $N=3$ and $Q_d=-\frac13$)
\begin{equation}\label{monster}
   \bm{\gamma}_Q^{(0)} = \left( \begin{matrix}
    \frac43\sin^2\theta -2 & 6 & 0 & 3 & 0 & 0 \\
    6 & \frac43\sin^2\theta -2 & \frac{416}{81} & \frac{70}{27} & 0 & 0 \\
    0 & 0 & \frac{32}{3} - \frac43\sin^2\theta & 0 & \frac83\sin2\theta
     & 0 \\
    0 & 0 & -\frac{32}{9} & \frac{28}{3} - \frac43\sin^2\theta & 0
     & \frac83\sin2\theta \\
    0 & 0 & \sin2\theta & 0 & \frac{43}{2} - \frac43\sin^2\theta & 0 \\
    0 & 0 & 0 & \sin2\theta & -\frac{32}{9}
     & \frac{121}{6} - \frac43\sin^2\theta 
   \end{matrix} \right) .
\end{equation}
The solution of the RG equation in this basis yields coefficients 
$(c_1,c_2,c_3,c_4,c_5,c_6)$ at a scale between $m_W$ and $\mg$. Their 
initial values at the electroweak scale are given by 
$(0,1,C_{7\gamma}(m_W),C_{8g}(m_W),0,0)$. The relevant $\beta$-function 
coefficient in this range is $\beta_0(5,1,1)$. From these solutions, we 
obtain the SM Wilson coefficients at the scale $\mu_{\tilde g}\sim\mg$ by 
means of the relations $C_{1,2}(\mu_{\tilde g})=c_{1,2}(\mu_{\tilde g})$ 
and
\begin{equation}
   C_{7\gamma}(\mu_{\tilde g}) = c_3(\mu_{\tilde g}) +
    \frac{\mg(\mu_{\tilde g})}{m_b(\mu_{\tilde g})}\,c_5(\mu_{\tilde g}) \,,
   \qquad 
   C_{8g}(\mu_{\tilde g}) = c_4(\mu_{\tilde g}) +
    \frac{\mg(\mu_{\tilde g})}{m_b(\mu_{\tilde g})}\,c_6(\mu_{\tilde g}) \,.
\end{equation}
The sign of the coefficients $c_{5,6}$ depends on the sign of the mixing 
angle $\theta$. At leading order, the running $b$-quark mass at the gluino 
scale is obtained from $m_b(\mu_{\tilde g})=m_b(m_b)\,
[\alpha_s(\mu_{\tilde g})/\alpha_s(m_b)]^{4/\beta_0(5,0,1)}$.

Once we have determined the SM contributions to the Wilson coefficients at 
the scale $\mu_{\tilde g}$, their evolution down to lower scales is governed 
by the well-known evolution equations of the SM. The corresponding $4\times 4$
anomalous dimension matrix coincides with the upper left $4\times 4$ corner
of the extended matrix in (\ref{monster}) evaluated at $\theta=0$. The
resulting formulae are more complicated than in the SM, because in our case
the coefficient $C_1(\mu_{\tilde g})$ does not vanish at the matching scale 
(whereas $C_1(m_W)=0$ for the standard evolution). 

\begin{table}
\footnotesize
\centerline{\parbox{15cm}{\caption{\label{tab:Ccoefs}
Results for the Wilson coefficients and the running $b$-quark mass for 
different values of $\mu$. Input parameters are $m_b(m_b)=4.2$\,GeV, 
$m_t(m_W)=174$\,GeV, $\mg(\mg)=15$\,GeV, and $\alpha_s(m_b)=0.21$. In the 
upper portion of the table the gluino is integrated out at $\mu=\mg$, in 
the lower portion at $\mu=2\mg$. If two signs are shown, the upper (lower) 
one refers to positive (negative) mixing angle $\theta$.}}}
\vspace*{0.2cm}
\begin{center}
\begin{tabular}{|c|c|cccc|ccc|}
\hline\hline
Scale & $m_b(\mu)$\,[GeV] & ${\cal C}_1$ & ${\cal C}_2$
 & ${\cal C}_7$ & ${\cal C}_8$ & $C_2$ & $C_{7\gamma}$ & $C_{8g}$ \\
\hline
$m_W$ & $3.17\mp 0.42$ & --- & --- & --- & ---
 & 1 & $-0.195$ & $-0.097$ \\
$\mg$ & 3.59 & 2 & 0 & $-0.222$ & $-0.833$
 & 1.040 & $-0.255\pm 0.028$ & $-0.124\pm 0.014$ \\
$m_b$ & 4.20 & 2.691 & 0.066 & $-0.264$ & $-0.804$
 & 1.104 & $-0.313\pm 0.023$ & $-0.143\pm 0.011$ \\
\hline
$m_W$ & $3.13\mp 0.25$ & --- & --- & --- & ---
 & 1 & $-0.195$ & $-0.097$ \\
$\mg$ & 3.59 & 2.274 & 0.025 & $-0.241$ & $-0.821$
 & 1.042 & $-0.255\pm 0.015$ & $-0.124\pm 0.008$ \\
$m_b$ & 4.20 & 3.064 & 0.104 & $-0.280$ & $-0.791$
 & 1.106 & $-0.313\pm 0.013$ & $-0.143\pm 0.006$ \\
\hline\hline
\end{tabular}
\end{center}
\end{table}

Table~\ref{tab:Ccoefs} shows the results for the Wilson coefficients at
different values of the renormalization scale. (The coefficient $C_1$
does not enter the $B\to X_s\gamma$ branching ratio and is omitted here.)
The values of $C_{7\gamma}$ and $C_{8g}$ depend on the sign of the mixing 
angle $\theta$, although this effect is numerically small. For comparison, 
the values obtained at $\mu=m_b$ in the SM (using $\alpha_s(m_Z)=0.118$) 
are $C_2\simeq 1.12$, $C_{7\gamma}\simeq -0.32$ and $C_{8g}\simeq -0.15$. 
They are very close to the values found in the presence of the light 
SUSY particles. In addition, there are the extra contributions proportional 
to the new coefficients ${\cal C}_i$. The second
column in the table shows the running $b$-quark mass at the various scales. 
Note that the value of $m_b$ above the gluino scale is very sensitive to 
the sign of $\theta$. The result $m_b(m_W)=2.75$\,GeV corresponding to
positive $\theta$ appears to be favored by the DELPHI measurement 
$m_b(m_W)=(2.67\pm 0.50)$\,GeV obtained from three-jet production 
of heavy quarks at LEP \cite{Abreu:1997ey}. However, here a similar 
comment as in our discussion of the running of $\alpha_s$ applies, namely 
that the DELPHI analysis implicitly assumes that $m_b$ runs as predicted in 
the SM. 

We are now ready to present our results for the $B\to X_s\gamma$ decay
rate, including both the SM and the new SUSY flavor-changing 
contributions. It is convenient to define new coefficients
\begin{eqnarray}
   C_7^{LR}(\mu) &=& C_{7\gamma}(\mu)
    - \frac{\sqrt2\pi\alpha_s(\mg)}{G_F\mg}\,
    \frac{\epsilon_{sb}^{LR}}{V_{ts}^* V_{tb}}\,
    \frac{{\cal C}_7(\mu)}{m_b(\mu)} \,, \nonumber\\
   C_7^{RL}(\mu) &=& - \frac{\sqrt2\pi\alpha_s(\mg)}{G_F\mg}\,
    \frac{\epsilon_{sb}^{RL}}{V_{ts}^* V_{tb}}\,
    \frac{{\cal C}_7(\mu)}{m_b(\mu)} \,,
\end{eqnarray}
and analogous coefficients $C_8^{LR}$ and $C_8^{RL}$.
These expressions exhibit the general features of our model
as described earlier. The SUSY contributions are enhanced relative to
the SM contributions by a large factor 
$\frac{\sqrt2\pi\alpha_s(\mg)}{G_F\mg m_b}\approx 10^3$, meaning that
the ratio of flavor-changing couplings, 
$\epsilon_{sb}^{AB}/(V_{ts}^* V_{tb})$, must be highly suppressed so as 
not to spoil the successful SM prediction for the branching ratio. The 
resulting leading-order expression for the $B\to X_s\gamma$ decay rate is
\begin{equation}\label{rate}
   \Gamma(B\to X_s\gamma)
   = \frac{G_F^2\alpha M_b^3 m_b^2(m_b)}{32\pi^4}\,
   |V_{ts}^* V_{tb}|^2 \Big[ |C_7^{LR}(\mu_b)|^2
   + |C_7^{RL}(\mu_b)|^2 \Big] \,,
\end{equation}
where $\mu_b\sim m_b$ is the renormalization scale. $M_b$ is a low-scale 
subtracted quark mass, which naturally enters the theoretical
description of inclusive $B$ decays once the pole mass is eliminated so
as to avoid bad higher-order perturbative behavior.

\begin{table}
\centerline{\parbox{15cm}{\caption{\label{tab:Rcoeff}
Results for the coefficients $B_0$ and $A_{1,2}$ for the SM (first row), 
and for the SUSY scenarios with positive (middle portion) and negative 
(lower portion) mixing angle $\theta$, for $E_\gamma>2$\,GeV. The quoted 
errors refer to the variations of the theoretical parameters within 
the ranges specified in the text. The renormalization scale is varied 
between 2.5 and 7.5\,GeV. Other input parameters are $V_{ts}^* V_{tb}=-0.04$,
$\alpha_s(m_b)=0.21$ for the SUSY scenario, and $\alpha_s(m_b)=0.225$ for 
the SM.}}}
\vspace*{0.2cm}
\begin{center}
\begin{tabular}{|c|cccc|}
\hline\hline
 & Default & $\Delta((m_b^{1S})^3\,m_b^2)$ & $\Delta(m_c/m_b)$
 & $\Delta\mu_b$ \\ 
\hline
$B_0^{\rm SM}$ & 3.44 & $\pm 0.21$ & $\mp 0.11$ & ~${}_{-0.16}^{+0.09}$~ \\
\hline
$B_0$ & 2.93 & $\pm 0.18$ & $\mp 0.09$ & ${}_{-0.13}^{+0.06}$ \\
$10^{-4}\,A_1$ & 3.60 & 0 & $\pm 0.05$ & ${}_{-0.02}^{+0.12}$ \\
$10^{-8}\,A_2$ & 3.12 & 0 & $\pm 0.10$ & ${}_{-0.12}^{+0.28}$ \\
\hline
$B_0$ & 3.68 & $\pm 0.23$ & $\mp 0.11$ & ${}_{-0.11}^{+0.04}$ \\
$10^{-4}\,A_1$ & 3.19 & 0 & $\pm 0.04$ & ${}_{-0.01}^{+0.06}$ \\
$10^{-8}\,A_2$ & 2.48 & 0 & $\pm 0.07$ & ${}_{-0.08}^{+0.18}$ \\
\hline\hline
\end{tabular}
\end{center}
\end{table}

It is well known that NLO corrections have a significant impact on the
$B\to X_s\gamma$ decay rate in the SM, which is largely due to NLO 
corrections to the matrix elements of the operators $Q_i$ in the effective 
weak Hamiltonian. (NLO corrections to the Wilson coefficient 
$C_{7\gamma}$ have a much smaller effect.) In order to capture the bulk of 
these corrections, we include the $O(\alpha_s)$ contributions to the 
matrix elements but neglect SUSY NLO corrections to the coefficients 
$C_7^{LR}$ and $C_7^{RL}$. 
We also neglect two-loop contributions to the matrix elements involving
$\tilde b$-squark loops. This is justified because of the relatively large 
mass of the $\tilde b$ squark, and because our two-loop anomalous dimension 
calculation has shown that there is very little mixing of the squark
operators into the dipole operators. At NLO our results become sensitive to
the precise definition of the mass parameter $M_b$, which we identify with
the so-called Upsilon mass \cite{Hoang:1998ng}, for which we take the 
value $m_b^{1S}=4.72\pm 0.06$\,GeV \cite{Neubert:2001ib}. (Up to a small 
nonperturbative contribution, 
$m_b^{1S}$ is one half of the mass of the $\Upsilon(1S)$ resonance.) We
also introduce a cutoff $E_\gamma^{\rm min}=\frac12(1-\delta)\,m_b^{1S}$ 
on the photon energy in the $B$-meson rest frame, which is required 
in the experimental analysis of radiative $B$ decays. We then obtain
\begin{equation}
   \mbox{Br}(B\to X_s\gamma)\big|_{E_\gamma>E_\gamma^{\rm min}}
   = \tau_B\,\frac{G_F^2\alpha (m_b^{1S})^3 m_b^2(m_b)}{32\pi^4}\,
   |V_{ts}^* V_{tb}|^2\,K_{\rm NLO}(\delta) \,,
\end{equation}
where $K_{\rm NLO}(\delta)$ is obtained from the formulae in 
\cite{Kagan:1998ym} by obvious modifications to include the effects 
of the new SUSY contributions, and by a change in some of the NLO terms 
due to the introduction of the Upsilon mass in (\ref{rate}) \cite{footnote}.
The dependence of the branching ratio on the SUSY flavor-changing couplings 
can be made explicit by writing
\begin{equation}\label{Aidef}
   \mbox{Br}(B\to X_s\gamma)\big|_{E_\gamma>E_\gamma^{\rm min}}
   = 10^{-4}\,B_0(\delta) \Big[
   1 + A_1(\delta)\,\text{Re}(\epsilon_{sb}^{LR})
   + A_2(\delta) \left( |\epsilon_{sb}^{LR}|^2 + |\epsilon_{sb}^{RL}|^2
   \right) \Big] \,.
\end{equation}
In Table~\ref{tab:Rcoeff}, we give results for the coefficients $B_0$ and 
$A_{1,2}$ including the dominant theoretical uncertainties. Following 
\cite{Gambino:2001ew}, we use a running charm-quark mass in the 
penguin-loop diagrams rather than the pole mass. This is justified, because 
the photon-energy cut imposed in the experimental analysis prevents the 
intermediate charm-quark propagators from being near their mass shell. 
Specifically, we work with the mass ratio $m_c(\mu)/m_b(\mu)$, where the 
running masses are obtained from $m_c(m_c)=(1.25\pm 0.10)$\,GeV and 
$m_b(m_b)=(4.20\pm 0.05)$\,GeV.

\begin{figure}
\begin{center}
\includegraphics[width=0.95\textwidth]{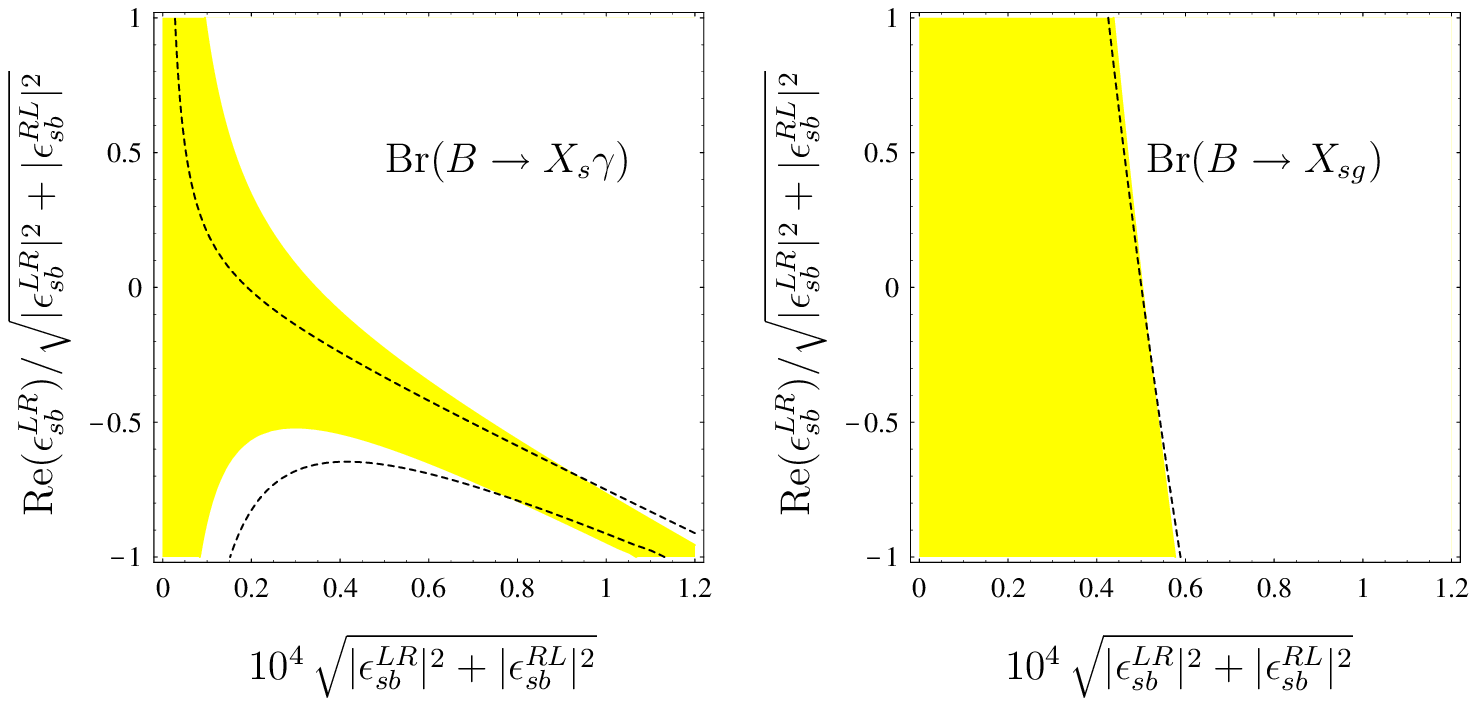}
\vspace*{-0.4cm}
\parbox{15cm}{\caption{\label{fig:Bsgamma}
Allowed regions (at 95\% c.l.) for the SUSY flavor-changing parameters 
obtained from the CLEO measurements of the $B\to X_s\gamma$ (left) and 
$B\to X_{sg}$ (right) branching ratios, using central values for all 
theory input parameters. The shaded regions correspond to the SUSY model 
with positive mixing angle $\theta$, the dashed lines refer to negative 
$\theta$.}}
\end{center}
\end{figure}

In the left-hand plot in Figure~\ref{fig:Bsgamma}, we confront our 
theoretical result for the $B\to X_s\gamma$ branching ratio with the CLEO 
measurement 
$\text{Br}(B\to X_s\gamma)=(3.06\pm 0.41\pm 0.26)\cdot 10^{-4}$ obtained 
for $E_\gamma>2$ GeV \cite{Chen:2001fj}. (This result actually corresponds 
to the sum of $B\to X_s\gamma$ and $B\to X_d\gamma$ decays. However, the
suppression of the exclusive $B\to\rho\gamma$ decay with respect to the
$B\to K^*\gamma$ mode implies that the dominant contribution to 
the inclusive decay must come from $b\to s\gamma$ transitions. In the
context of our model, it follows that the couplings $\epsilon_{db}^{AB}$
must obey even tighter constraints than the $\epsilon_{sb}^{AB}$.) It
follows that the maximum allowed values of the parameters 
$\epsilon_{sb}^{AB}$ are $10^{-4}$, as is already obvious from the 
magnitude of the coefficients $A_{1,2}$ in Table~\ref{tab:Rcoeff}. However, 
values larger than $5\cdot 10^{-5}$ would require a fine-tuning of the 
phase of $\epsilon_{sb}^{LR}$ and are thus somewhat unnatural. Note
that the ratio shown on the vertical axis in the plot is bound to lie 
between 1 and $-1$, and in the limit $\epsilon_{sb}^{RL}=0$ corresponds to 
$\cos\vartheta_{LR}$, where $\vartheta_{LR}$ denotes the CP-violating phase 
of $\epsilon_{sb}^{LR}$.

The right-hand plot in the figure shows a similar constraint arising from 
the inclusive charmless decay $B\to X_{sg}$. At leading order, the decay 
rate for this process is obtained from (\ref{rate}) by the replacements 
$\alpha\to\frac43\,\alpha_s(\mu_b)$ and $C_7^{AB}\to C_8^{AB}$. The allowed 
region corresponds to the CLEO upper bound of 8.2\% (at 95\% c.l.) for the 
$B\to X_{sg}$ branching ratio \cite{Coan:1997ye}. There are, however, 
potentially large theoretical uncertainties in this result, because we 
neglect NLO corrections to the branching ratio. We therefore refrain from 
combining the two plots to reduce the allowed parameter space.

In the SM, the direct CP asymmetry in the inclusive decay $B\to X_s\gamma$
is very small, below 1\% in magnitude \cite{Kagan:1998bh}. In the SUSY 
scenario, on the other hand, the phase of the coupling $\epsilon_{sb}^{LR}$ 
could lead to a large asymmetry. In the approximation where one neglects 
the SUSY contributions to the CP-averaged decay rate in (\ref{rate}), which 
is justified in view of the good agreement of the SM prediction with the 
data, the formulae in \cite{Kagan:1998bh} yield the prediction
$A_{\rm CP}\approx -50\%\times 10^4\,\text{Im}(\epsilon_{sb}^{LR})$, 
where we have neglected the small contribution from the charm-quark loops 
and the yet smaller contribution from $\tilde b$-squark loops. (We use the 
standard phase convention where $\lambda_t=V_{ts}^*\,V_{tb}$ is real. In 
general, the CP asymmetry depends on 
$\text{Im}(\epsilon_{sb}^{LR}/\lambda_t)$.) It follows that even within 
the very restrictive bounds shown in Figure~\ref{fig:Bsgamma} there can 
be a potentially large contribution to the CP asymmetry, which would 
provide a striking manifestation of physics beyond the SM. In fact, the 
CLEO bounds $-27\%<A_{\rm CP}<+10\%$ (at 90\% c.l.) \cite{Coan:2000pu} 
imply that 
\begin{equation}
   -2\cdot 10^{-5} < \text{Im}(\epsilon_{sb}^{LR}) < 5\cdot 10^{-5} \,,
\end{equation}
placing another tight constraint on the flavor-changing coupling
$\epsilon_{sb}^{LR}$.

\section{Conclusions}

New supersymmetric contributions to $b$-quark production at hadron colliders 
can account for the long-standing discrepancy between the measured cross 
sections and QCD predictions if there is a light $\tilde b$ squark with 
mass in the range 2--5.5\,GeV, accompanied by a somewhat heavier gluino 
\cite{Berger:2000mp}. In this Letter, we have explored the phenomenology of 
rare $B$ decays in such a scenario and have found tight bounds 
on the flavor-changing parameters controlling supersymmetric 
contributions to $b\to s$ and $b\to d$ FCNC transitions. The most 
restrictive constraints arise from virtual effects of light 
$\tilde b$ squarks in $B\to X_s\gamma$ decays. We have analysed this process
by constructing a low-energy effective Hamiltonian, in which the gluinos are
integrated out, while the $\tilde b$ squarks remain as dynamical degrees of 
freedom. We find that the flavor-changing couplings $\epsilon_{sb}^{RL}$ and 
$\epsilon_{sb}^{LR}$ must be of order few times $10^{-5}$ or less. (Even 
tighter constraints hold for the analogous $b\to d$ couplings.) This implies 
that certain off-diagonal entries of the down-squark mass matrix must be 
suppressed by a similar factor compared to the generic squark-mass squared. 

Even with such tight constraints on the couplings, this model allows for
interesting and novel New Physics effects in weak decays of $B$ mesons 
and beauty baryons. As an example, we have discussed the direct CP 
asymmetry in $B\to X_s\gamma$ decays, which could be enhanced with respect
to its Standard Model value by an order of magnitude. Other possible 
effects include an enhanced $B\to X_{sg}$ decay rate. We have not 
considered here the possibility of $\tilde b$-squark pair-production, which
would be kinematically allowed for very light squark masses. The new decay
modes $b\to s\tilde b\tilde b^*$ and $b\to\bar s\tilde b\tilde b$ would 
affect the decay widths of $B$ mesons and $\Lambda_b$ baryons differently, 
and hence might explain the anomaly of the low $\Lambda_b$ lifetime. 
We will report on this interesting possibility elsewhere.

\vspace{0.2cm}  
{\it Acknowledgments:\/} 
S.B.\ and M.N.\ are supported by the National Science Foundation under 
Grant PHY-0098631. T.B.\ is supported by the Department of Energy under
Grant DE-AC03-76SF00515, and A.K.\ under Grant DE-FG02-84ER40153.


\begin{thebibliography}{99}
\bibitem{footnote}
Specifically, the expression for the quantity $k_{77}(\delta,\mu_b)$ in 
eq.~(13) of \protect\cite{Kagan:1998ym} must be replaced by 
{\small
\[
   k_{77}(\delta,\mu_b) = S(\delta)\,\left\{ 1
   + \frac{\alpha_s(\mu_b)}{2\pi} \left( r_7
   + \gamma_{77}\ln\frac{m_b}{\mu_b}\right) + \frac23\,\alpha_s^2(\mu_b)
   + \frac{\lambda_1-9\lambda_2}{2(m_b^{1S})^2} \right\}
   + \frac{\alpha_s(\mu_b)}{\pi}\,f_{77}(\delta) \,,
\]}\noindent
where $-\lambda_1\approx(0.25\pm 0.15)\,$GeV$^2$ and 
$\lambda_2\approx 0.12$\,GeV$^2$ are hadronic parameters.
\end{thebibliography}
\end{document}